\documentclass[12pt]{iopart}
\usepackage{graphicx}
\usepackage{color}

\renewcommand{\vec}[1]{\mbox{\boldmath $#1$}}

\begin{document}

\title
[Branching ratios for invisible dinucleon decays in $^{16}$O]
{
Branching ratios for deexcitation processes of daughter nuclei following 
invisible dinucleon decays in $^{16}$O}

\author{K.~Hagino$^{1,2}$ and M.~Nirkko$^3$}

\address{$^1$ 
Department of Physics, Tohoku University, Sendai 980-8578, Japan}
\address{$^2$ 
Research Center for Electron Photon Science, Tohoku University, 1-2-1 Mikamine, Sendai 982-0826, Japan}
\address{$^3$ 
Department of Physics and Astronomy, University of Sussex, Falmer, Brighton BN1 9RH, United Kingdom}


\begin{abstract}
Various theories beyond the standard model of particle physics predict the existence of baryon number violating processes resulting in nucleon decay.
When occurring within an atomic nucleus, such a decay will be followed by secondary decays of the daughter nucleus unless its ground state is directly populated.
In this paper, we estimate branching ratios for processes associated with dinucleon decays of the $^{16}$O nucleus.
To this end, we use a simple shell model for the ground state of $^{16}$O.
For decays from the 1$s_{1/2}$ configuration, which result in highly excited states in the daughter nucleus, we employ a statistical model with the Hauser-Feshbach theory.
Our analysis indicates that the branching ratio for gamma-ray emission in the energy range between 5 and 9~MeV, which is relevant to low-threshold water Cherenkov experiments such as SNO+, is 4.53\%, 35.7\%, and 20.2\% for the $nn$, $pp$, and $pn$ decays in $^{16}$O, respectively.
In particular, emission of 6.09~MeV and 7.01~MeV gamma-rays from $^{14}$C, and 6.45 MeV and 7.03 MeV gamma-rays from $^{14}$N, have branching ratios of as large as 10.9\%, 20.1\%, 7.73\% and 8.90\%, respectively. 
\end{abstract}

%
\vspace{2pc}
\submitto{\JPG}
%
\maketitle
%
%

\section{Introduction}

While a proton is stable in the standard model of particle physics, grand unified theories (GUTs) predict that it decays by violating baryon number conservation \cite{GUT}.
Similar decays are predicted for bound neutrons also.
In fact, the Particle Data Group lists 73 possible decay modes, both for one-nucleon and two-nucleon decays \cite{PDG}.
A recent measurement by the Super-Kamiokande collaboration sets the lower limit on the proton lifetime at 1.6$\times$ 10$^{34}$ and 7.7$\times$ 10$^{33}$ years at 90\% confidence level for the $p\to e^+\pi^0$ and $p\to \mu^+\pi^0$ modes, respectively \cite{Abe17}.
See also Refs.~\cite{Abe17-2,Litos14,Takhistov15,Gustafson15} for other nucleon decay searches.

Among the possible decay modes of nucleons, there are certain modes in which the final decay products are almost undetectable, such as $n\to \nu_e\nu_e\bar{\nu}_e$ and $nn\to \nu_e\bar{\nu}_e$ \cite{PDG}.
These are commonly referred to as invisible decays.
When the nucleon decay takes place in a nucleus, the resulting daughter nucleus will generally be in an excited state.
Such a state will de-excite to a lower energy state by emitting gamma-rays or nucleons.
Even though the primary decay products are invisible, invisible nucleon decays can therefore still be detected by measuring the secondary decays of the daughter nucleus \cite{Totsuka86,Ejiri93,Kamyshkov03}.
This strategy has been pursued in the past \cite{Suzuki93,Back03,Ahmed04,Araki06}, but the lower limits on the lifetime for the invisible decay modes are typically a few orders of magnitudes lower than that for the visible modes involving charged particles, which have higher energies and are therefore easier to detect.
For instance, the lower limits of lifetimes for the $n\to$ invisible and the $nn\to$ invisible modes have been reported to be 5.8$\times$ 10$^{29}$ and 1.4$\times$ 10$^{30}$ years at 90\% confidence level, respectively \cite{Araki06}.

To improve the current limits for invisible nucleon decay, low-background water Cherenkov detectors may be used.
Notably, the SNO+ experiment is conducting a search for nucleon decay in $^{16}$O in its initial phase using 900 tons of ultra-pure water, with $5.4-9$~MeV being the favourable energy region where backgrounds are expected to be low \cite{SNO+,SNO+2,Caden17}.
Thanks to recent upgrades to the detector electronics, a search using Super-Kamiokande may also be feasible \cite{Sekiya17}.

In order to extract a limit on the nucleon lifetimes from such measurements, it is crucial to estimate the branching ratios for the secondary decays of the daughter nucleus, since experiments can only measure the gamma-ray yields/limits.
While detailed studies on the branching ratios for the single nucleon decay modes in $^{16}$O exist \cite{Ejiri93,Kamyshkov03}, no such study has been carried out for dinucleon decays in $^{16}$O to the best of our knowledge.

Thus, the aim of this paper is to calculate the branching ratios for the secondary decays of daughter nuclei generated by dinucleon decays in $^{16}$O.
Such study is of considerable importance given that new searches for nucleon decay in water are currently ongoing \cite{Caden17}.
To this end, we follow a similar approach as that in Ref.~\cite{Kamyshkov03}.
That is, for secondary decays associated with the dinucleon decay from the 1$s_{1/2}$ orbit in $^{16}$O, we estimate the branching ratios using a statistical model code.
On the other hand, for secondary decays for the dinucleon decay from the $1p_{1/2}$ and/or the $1p_{3/2}$ orbits, we use the experimentally known decay properties of low-lying states in the daughter nuclei.

The paper is organized as follows:
In Sec.~II, we discuss the structure of the $^{16}$O nucleus and present a theoretical formula for a population probability of a state in a daughter nucleus following the two-nucleon decays.
In Sec.~III, we apply the formula to the $nn$, $pp$, and $pn$ decays of $^{16}$O and discuss the branching ratios for the secondary decays of the daughter nuclei.
We then summarize the paper in Sec.~IV.


\section{The structure of $^{16}$O and population probabilities for final states in the daughter nuclei}

\begin{figure}[t]
\begin{center}
\includegraphics[clip,width=8cm]{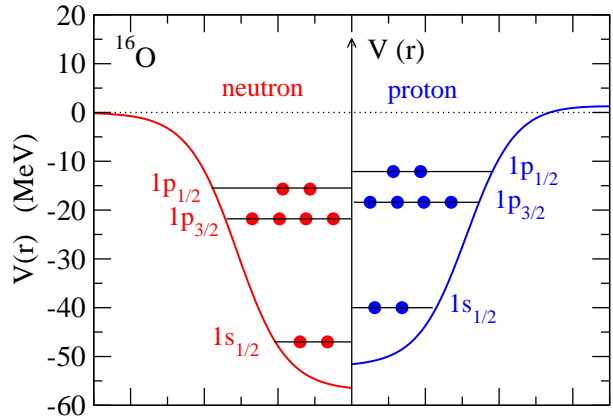}
\caption{A schematic illustration of single-particle levels in a mean-field potential $V(r)$ for the $^{16}$O nucleus.}
\label{fig1}
\end{center}
\end{figure}

$^{16}$O is a well known double-magic nucleus, and it is reasonable to assume that its ground state can be described with a simple shell model based on the mean field approximation.
In this approximation, 8 neutrons and 8 protons in $^{16}$O occupy single-particle levels up to the $N=8$ and the $Z=8$ shell gaps, respectively.
That is, 4 nucleons (2 neutrons and 2 protons) are in the 1$s_{1/2}$ state, 8 nucleons in the 1$p_{3/2}$ state, and 4 nucleons in the 1$p_{1/2}$ state, as schematically illustrated in Fig.~\ref{fig1}.
The observed single-particle energies $\epsilon$ \cite{BM69} are summarized in Table \ref{tab0}.
Here, the energies for the 1$p_{1/2}$ states are estimated from the one-nucleon separation energies of $^{16}$O, while the energies of the 1$p_{3/2}$ states are from the excitation energies of the $3/2^-$ states in $^{15}$O and $^{15}$N.
The energies for the 1$s_{1/2}$ states, on the other hand, are deduced from the $(p,2p)$ experiment \cite{BM69,Tyren66}.

\begin{table}[bht]
\caption{Empirical single-particle energies $\epsilon$ for neutrons and protons in $^{16}$O \cite{BM69}.
See also Refs.~\cite{CS72,PVF78,BVM74}.}
\label{tab0}
\begin{center}
\begin{tabular}{c|cc}
\hline
\hline
state & $\epsilon_n$ (MeV) & $\epsilon_p$ (MeV) \\
\hline
1$p_{1/2}$ & $-15.7$ & $-12.1$ \\
1$p_{3/2}$ & $-21.8$ & $-18.4$ \\
1$s_{1/2}$ & $-47.0$ & $-40.0\pm 8$ \\
\hline
\hline
\end{tabular}
\end{center}
\end{table}

The wave function for the ground state of $^{16}$O is then given by, %
\begin{equation}
|^{16}{\rm O}\rangle = \prod_{l=0,1}\prod_{j=l\pm 1/2}\prod_{m=-j}^j\prod_{t_z=p,n}
a_{jlmt_z}^\dagger|0\rangle, \label{wf16o}
\end{equation}
where $m$ is the $z$-component of the single-particle angular momentum $j$, $l$ is the single-particle orbital angular momentum, and $t_z=p,n$ is the $z$-component of isospin for each nucleon.
Here, $a_{jlmt_z}^\dagger$ is the creation operator of a nucleon with a specific quantum number $(j,l,m,t_z)$, and $|0\rangle$ is the vacuum state.
For simplicity of the notation, we have dropped the radial quantum number in the creation operator.

The population probability for a specific state $|f\rangle$ in a daughter nucleus following a dinucleon decay is given by, %
\begin{equation}
P_{\rm pop}(f)\propto
\sum_{s_z,s_z'}\sum_{t_z,t_z'}\int d\vec{r}
\left|\left\langle f\left|a_{\vec{r}s_zt_z}a_{\vec{r}s'_zt'_z}\right
|^{16}{\rm O}\right\rangle\right|^2, \label{pop_prob}
\end{equation}
where $a_{\vec{r}s_zt_z}$ is the annihilation operator for a nucleon at the position $\vec{r}$ with the isospin $t_z$ and the spin $s_z$.
Since baryon non-conserving dinucleon decays take place only when two nucleons are at a very short distance, we consider an elimination of two nucleons at the same position, although the short range correlation may also play a role \cite{Horiuchi11,Cirigliano18}.

In the following, we consider a pure configuration for the final state, $|f\rangle$, which is constructed by eliminating two nucleons from the single-particle levels of $(jl)$ and $(j'l')$ as, %
\begin{eqnarray}
|f\rangle&=&|jlj'l'; IM\rangle
={\cal N}[a_{jl}a_{j'l'}]^{(IM;TT_z)}|^{16}{\rm O}\rangle,   \\
&=&{\cal N}\sum_{m,m'}\sum_{t_z,t_z'}
\langle jmj'm'|IM\rangle
\left\langle\left. \frac{1}{2}t_z \frac{1}{2}t_z'\right|TT_z\right\rangle 
a_{jlmt_z}a_{j'l'm't_z'}|^{16}{\rm O}\rangle, \label{finalwf}
\end{eqnarray}
where $I$ is the angular momentum of the final state and $M$ is its $z$-component, while $T$ and $T_z$ are the total isospin and its $z$-component, respectively.
The normalization factor ${\cal N}$ is given by ${\cal N}=[1-(-1)^{I+T}\delta_{jl,j'l'}]^{-1/2}$.
As we show in the Appendix, for $T=0$, Eq.~(\ref{pop_prob}) is evaluated as
\begin{eqnarray}
P_{\rm pop}(IM)&=&I_r{\cal N}^2\,\frac{4(2j+1)(2j'+1)}{(2I+1)} \nonumber \\
&\times& \,\left(
\left\langle\left. j\frac{1}{2}j'\frac{1}{2}\right|I1\right\rangle^2
+
\left\langle\left. j\frac{1}{2}j'-\frac{1}{2}\right|I0\right\rangle^2
\,\delta_{l+l'-I,odd}\right), 
\label{popP0}
\end{eqnarray}
where $I_r$ is the radial integral of the product of single-particle wave functions given by Eq.~(\ref{I_r}).
For $T=1$, on the other hand, Eq.~(\ref{pop_prob}) is evaluated as 
\begin{equation}
 P_{\rm pop}(IM)=I_r{\cal N}^2\,\frac{4(2j+1)(2j'+1)}{(2I+1)} 
\left\langle\left. j\frac{1}{2}j'-\frac{1}{2}\right|I0\right\rangle^2
\,\delta_{l+l'-I,even}.
\label{popP1}
\end{equation}
Notice that $\langle j~1/2~j'~-1/2|I0\rangle^2$ is $\delta_{j,j'}/(2j+1)$ for $I=0$, and thus $P_{\rm pop}(IM)$ is proportional to the number of 0$^+$ pair in the single-particle orbit, $(2j+1)/2$.
However, the formula is somewhat more complicated than this intuitive picture for $I \neq 0$.

The total population probability for the specific state is then given as %
\begin{equation}
P_{\rm pop}(I)=\sum_MP_{\rm pop}(IM)=(2I+1)P_{\rm pop}(IM).
\label{Ptot}
\end{equation}

Notice that for the wave function with fully occupied single-particle levels, Eq.~(\ref{wf16o}), the population probabilities given by Eqs.~(\ref{popP0}) and (\ref{popP1}) involve only the recouplings of the single-particle angular momenta and the isospins, and have nothing to do with the properties of nucleon-nucleon interaction.
For instance, the same probabilities would be obtained even if the tensor interaction was completely absent, even though the tensor interaction usually stabilizes a $pn$-pair in the $T=0$ channel.
In this case, the effect of nucleon-nucleon interaction appears only through the structure of the daughter nuclei, particularly, the excitation energies.
This is in a marked contrast to a nucleus with one nucleon pair outside a closed shell nucleus, such as $^{18}$F, whose low-lying structure can be understood in terms of $^{16}{\rm O+p+n}$.
In $^{18}$F, the tensor interaction between the valence proton and neutron stabilizes the $T=0$ configuration and the ground state of $^{18}$F is 1$^+$, whereas the $T=1$ configuration appears at an excited state, that is, the first 0$^+$ state at 1.04 MeV \cite{Tanimura14}.
In this case, the invisible $pn$ decay would take place exclusively from the $T=0$ channel.
In the case of $^{16}$O, on the other hand, both $T=0$ and $T=1$ channels contribute, since the wave function of $^{16}$O already contains both of these components. 

Likewise, $nn$ and $pp$ pairs with $I\neq 0$ will also contribute to the final populations.
Even though the spatial overlap between single-particle wave functions are maximized for an $I=0$ pair and thus the ground state usually takes $I^\pi=0^+$ in even-even nuclei \cite{RS80}, the reduction in the spatial overlap for $I\neq 0$ is well compensated by the factor $(2I+1)$ in Eq.~(\ref{Ptot}).
As a consequence, the population probability for an $I\neq0$ pair is comparable to, or can even be larger than, that for an $I=0$ pair (see Tables \ref{tab:Ppop14O},\ref{tab:Ppop14C}, and \ref{tab:Ppop14N} below).


\section{Branching ratios for secondary decays of the daughter nucleus}


\subsection{Two-neutron decay in $^{16}$O}
We now apply the formalism presented in the previous section to the dinucleon decays of $^{16}$O and discuss the branching ratios for the secondary decays of the daughter nuclei.
We first consider an invisible dineutron decay, such as $nn\to 2\nu$.
In this paper, for simplicity, we ignore the pairing correlation among neutrons.
In this naive shell model, the ground state of $^{14}$O is populated when the neutron pair in $1p_{1/2}$ is removed, while the second 0$^+$ state of $^{14}$O is generated when a neutron pair in $1p_{3/2}$ disappears.

To justify this simplification, we have estimated the effect of pairing correlation using the hole-hole Tamm-Dancoff approximation \cite{RS80}, which is similar to a three-body model with a core+2 valence neutrons \cite{BE91,HS05}.
Using a Woods-Saxon single-particle potential to reproduce the energies of the $1p_{1/2}$ and the $1p_{3/2}$ states, together with a simple contact interaction $v_{\rm pair}(\vec{r},\vec{r}')=-g\delta(\vec{r}-\vec{r}')$ between the hole states, where the strength $g$ is determined so that the empirical two-neutron separation energy of $^{16}$O, $S_{2n}(^{16}$O$)=28.89$~MeV is reproduced, we have found that the ground state of $^{14}$O is composed of the $(1p_{1/2})^{-2}$ configuration with 94.3\%.
It is therefore a reasonable approximation to ignore the pairing correlation in discussing the branching ratio for the dineutron decay of $^{16}$O.
We expect a similar amount of mixing for the first and the second 0$^+$ states in $^{14}$C and $^{14}$N.
For the first and the second 1$^+$ states in $^{14}$N, we have confirmed that the mixing is even smaller, with 99.2\% for the $(1p_{1/2})^{-2}$ configuration in the first 1$^+$ state. 
We thus neglect the pairing correlation also for the diproton decay and the $pn$ decay to be discussed in the next subsections.
In the following, we assign the shell model configurations to the observed states based on the excitation energy of each configuration.

In addition to the two $0^+$ states in $^{14}$O, one also needs to take into account $I^\pi=2^+$ pairs with the $|(1p_{1/2})^{-1}(1p_{3/2})^{-1}\rangle$ configuration, as well as a 2$^+$ pair formed with two neutron holes in the $1p_{3/2}$ orbits.
Again, we neglect the pairing correlations and assign the former and the latter states to the first and the second 2$^+$ states in $^{14}$O at 6.59 and 7.77 MeV, respectively.

The population probabilities for these states in $^{14}$O, estimated with Eq.~(\ref{popP1}), are summarized in Table \ref{tab:Ppop14O}.
To obtain the single-particle wave functions in $I_r$,  we employ a Woods-Saxon potential, for which the radius parameter and the surface diffuseness parameter are taken to be $R_0=1.235\times 16^{1/3}$ fm and $a=$ 0.67 fm, respectively, while the depth parameter is adjusted for each single-particle level to reproduce the empirical neutron single-particle energies of $^{16}$O.
The population probabilities are then normalized, together with the probabilities for the other final states with neutron holes in the $1s_{1/2}$ level.

\begin{table}[bht]
\caption{The population probabilities, $P_{\rm pop}$, for the final states in the daughter nucleus $^{14}$O  for the $nn$ decay of $^{16}$O.
$T$ and $I^\pi$ are the total isospin and the spin-parity for each state, respectively.
$E^*$ is the excitation energy, where B.W. indicates that the energy is distributed according to the Breit-Wigner function (see Eq.~(\ref{breit-wigner})). 
The configuration and the dominant decay mode for each state are also shown.}
\label{tab:Ppop14O}
\begin{center}
\begin{tabular}{cc|c|c|c|c}
\hline
\hline
$T$ & $I^\pi$ & configuration & $E^*$ (MeV) & $P_{\rm pop}$ & decay \\
\hline
1 & 0$^+$ & (1$p_{1/2}$)$^{-2}$ & 0 & 0.0466 & - \\
1 & 0$^+$ & (1$p_{3/2}$)$^{-2}$ & 5.92 & 0.109 & p emission\\
1 & 2$^+$ & (1$p_{1/2}$)$^{-1}$(1$p_{3/2}$)$^{-1}$ & 6.59 & 0.201 & p emission\\
1 & 2$^+$ & (1$p_{3/2}$)$^{-2}$ & 7.77 & 0.109 & p emission\\
1 & 1$^-$ & (1$p_{1/2}$)$^{-1}$(1$s_{1/2}$)$^{-1}$ & B.W. & 0.119 & statistical\\
1 & 1$^-$ & (1$p_{3/2}$)$^{-1}$(1$s_{1/2}$)$^{-1}$ & B.W. & 0.270 & statistical \\
1 & 0$^+$ & (1$s_{1/2}$)$^{-2}$ & B.W. & 0.146 & statistical \\
\hline
\hline
\end{tabular}
\end{center}
\end{table}

\begin{figure}[t]
\begin{center}
\includegraphics[clip,width=7cm]{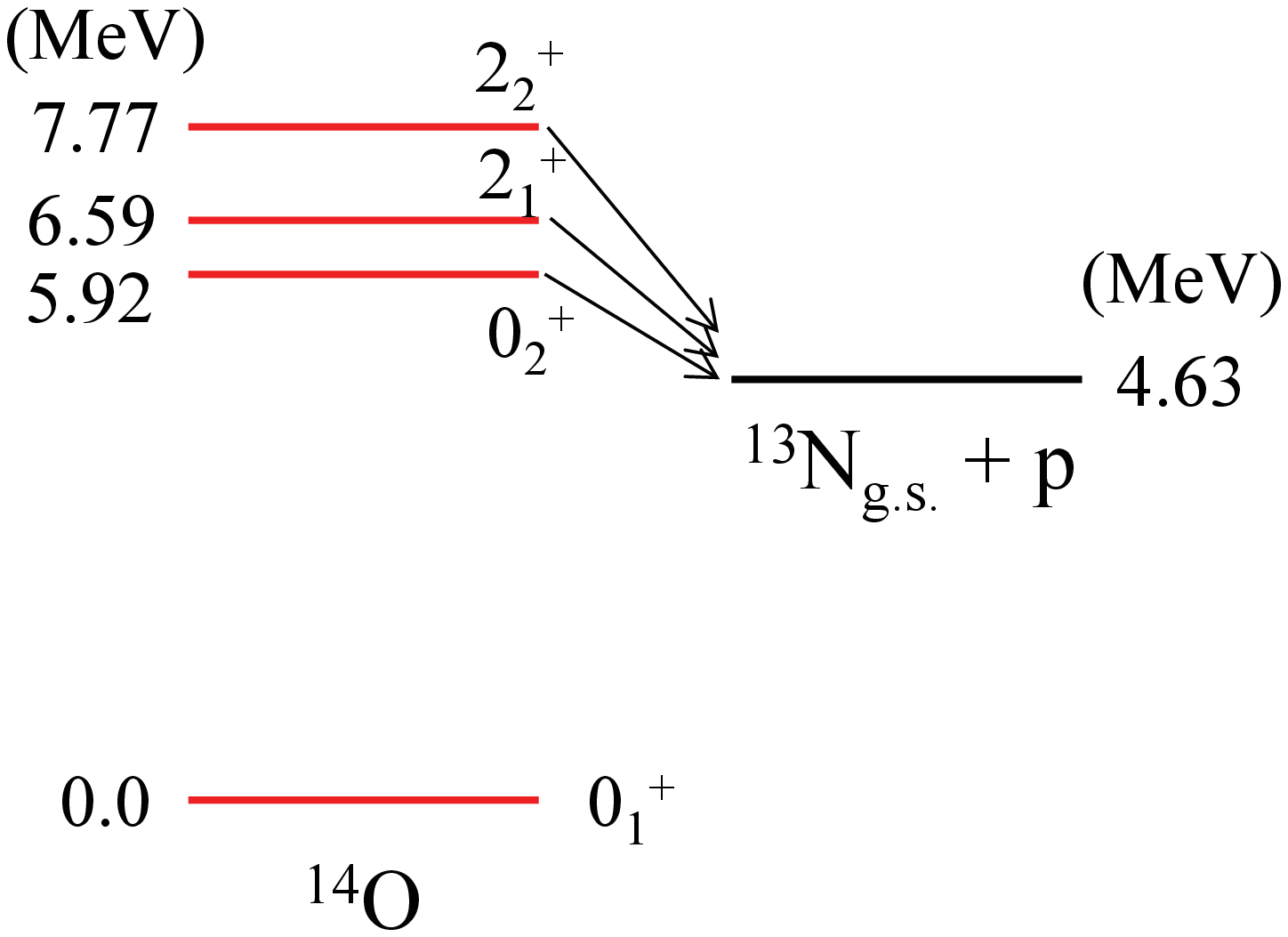}
\caption{A decay scheme for the excited stated in $^{14}$O originated from the dineutron decay from the 1$p_{1/2}$ and 1$p_{3/2}$ orbits in $^{16}$O.}
\label{fig2}
\end{center}
\end{figure}

Notice that the one-proton separation energy of $^{14}$O is $S_p(^{14}$O) = 4.63~MeV, therefore the second 0$^+$ state at 5.92~MeV, the first 2$^+$ state at 6.59 MeV, and the second 2$^+$ state at 7.77 MeV decay by emitting a proton to $^{13}$N(g.s.) with 100\% probability \cite{nndc}, as illustrated in Fig.~\ref{fig2}.
The branching ratios $\mathcal{B}$ for the ground state of $^{14}$O and $^{13}$N following the dineutron decays from the 1$p_{3/2}$ and 1$p_{1/2}$ orbits are given by $\mathcal{B}[^{14}{\rm O}(0_1^+)]=0.0466$ and $\mathcal{B}[^{14}{\rm O}(0_2^+)]=0.419$, respectively.
Notice that, in estimating those branching ratios, we neglect the final state kinematics, such as the difference in the phase space factor of the undetected particles produced in the dinucleon decay process.
However, such effects should be negligibly small as long as the undetected particles are light enough.

To these branching ratios, one also needs to add contributions of a dineutron decay from the 1$s_{1/2}$ orbit.
The removal of a neutron from the 1$s_{1/2}$ orbit results in a fragmentation of the strength in a wide energy region \cite{Ejiri93}.
Following Ref.~\cite{Kamyshkov03}, we assume that the strength is distributed according to the Breit-Wigner (B.W.) function %
\begin{equation}
f(I,E^*)=P_{\rm pop}(I)\cdot\frac{2}{\pi\Gamma}\, \frac{\Gamma^2/4}{(E^*-E_{0I}^*)^2+\Gamma^2/4},
\label{breit-wigner}
\end{equation}
with the width of $\Gamma=7$~MeV.
Here, we have taken into account the population probability for each state, $P_{\rm pop}(I)$.
The centroid energy, $E_{0I}^*$, is estimated as $E_{0I}^*=-\epsilon_n(jl)-\epsilon_n(j'l')-S_{2n}(^{16}{\rm O})$ for each configuration (see Table \ref{tab:Ppop14O}).

Such highly excited states of $^{14}$O decay by emitting a number of particles, such as neutrons, protons, deuterons, tritons, $^3$He, and $\alpha$-particles, as well as gamma-rays.
We evaluate these decays using the statistical model provided by the {\tt TALYS} software \cite{TALYS} with the default parameter set.
This code uses the Hauser-Feshbach theory \cite{Frobrich96} with the Gilbert-Cameron level density \cite{GC65} and the optical potentials of Koning and Delaroche \cite{KD03}.

\begin{figure}[t]
\begin{center}
\includegraphics[clip,width=8cm]{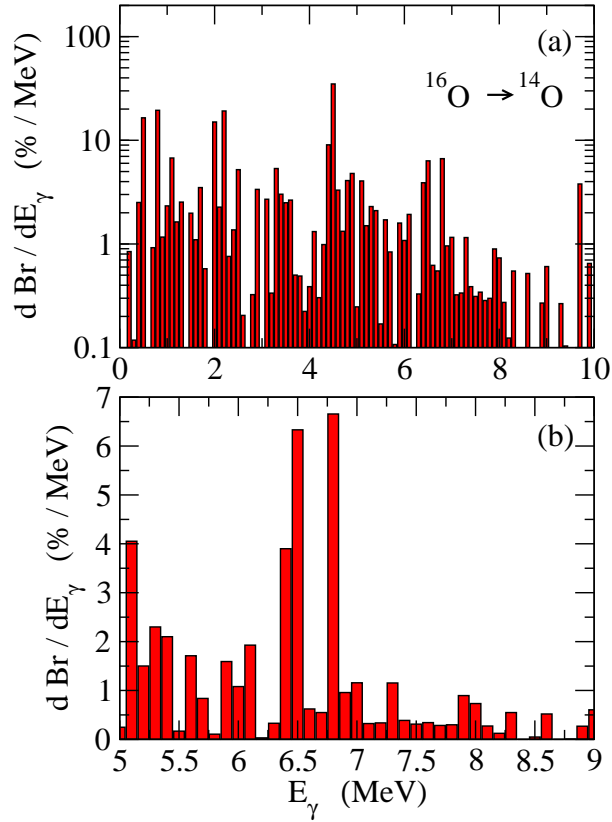}
\caption{
(a) Spectrum of gamma-rays from secondary decays of $^{14}$O originated from the dineutron decay of $^{16}$O as a function of the gamma-ray energy $E_\gamma$.
The width of the energy bins is 0.1~MeV.
The accumulated branching ratio in the region not shown in the figure amounts to 2.71\%.
(b) Expanded view of the upper panel in the region of 5~MeV~$\leq E_\gamma \leq 9$~MeV, which is particularly relevant to current experiments.
}
\label{fig3}
\end{center}
\end{figure}

\begin{table}[bht]
\caption{Branching ratios $\mathcal{B}$ for the final decay products for the dineutron decays of $^{16}$O, in which g.s. stands for the ground state of each nucleus.}
\label{tab1}
\begin{center}
\begin{tabular}{cc|cc}
\hline
\hline
Nucleus & $\mathcal{B}$ (\%) & Nucleus & $\mathcal{B}$ (\%) \\
\hline
$^{14}$O (g.s.) & 4.95 & $^{13}$N (g.s.) & 42.1 \\
$^{12}$C (g.s.) & 6.19 & $^{11}$C (g.s.) & 5.87 \\
$^{11}$B (g.s.) & 5.87 & $^{10}$B (g.s.) & 2.77 \\
$^{9}$B (g.s.) & 4.73 & $^{9}$Be (g.s.)  & 2.24 \\
$^{8}$Be (g.s.) & 20.46 & $^{7}$Be (g.s.) & 2.49 \\
$^{6}$Li (g.s.) & 2.55 & &  \\
\hline
\hline
\end{tabular}
\end{center}
\end{table}

\begin{table}[bht]
\caption{
Branching ratios $\mathcal{B}$ for the dominant discrete gamma-ray emissions  for the dineutron decays of $^{16}$O.
The number in the parenthesis denotes the energy of each state.}
\label{tab2}
\begin{center}
\begin{tabular}{cccc}
\hline
\hline
Nucleus & Transition & $E_\gamma$ (MeV) & $\mathcal{B}$ (\%) \\
\hline
$^{12}$C & 2$_1^+$ (4.44) $\to$ 0$_1^+$ (0.0) & 4.44 & 3.01 \\
$^{11}$C & 1/2$_1^-$ (2.00) $\to$ 3/2$_1^-$ (0.0) & 2.00 & 1.45 \\
$^{11}$B & 1/2$_1^-$ (2.12) $\to$ 3/2$_1^-$ (0.0) & 2.12 & 1.37 \\
$^{10}$B & 1$_1^+$ (0.72) $\to$ 3$_1^+$ (0.0) & 0.72 & 1.79 \\
$^{7}$Be & 1/2$_1^-$ (0.429) $\to$ 3/2$_1^-$ (0.0) & 0.429 & 0.852 \\
\hline
\hline
\end{tabular}
\end{center}
\end{table}

The results of the {\tt TALYS} calculation are summarized in Tables \ref{tab1} and \ref{tab2} for the branching ratios of the final decay products and those of the dominant discrete gamma-ray emissions, respectively.
For the former, we have included the contributions from the decays from the 1$p_{1/2}$ and the 1$p_{3/2}$ orbits.
For the latter, we have removed those 
from unbound states, for which the gamma-ray branching is 
much smaller than that for particle emissions.
The gamma-ray spectrum is shown in Fig.~\ref{fig3} as a function of energy $E_\gamma$.

Since the dineutron decays from the $1p_{1/2}$ and the $1p_{3/2}$ configurations of $^{16}$O result in the direct population of the ground state of $^{14}$O and $^{13}$N without gamma-ray emission (see Fig.~\ref{fig2} and Table \ref{tab1}), the gamma-ray branch originates entirely from the decay from the $1s_{1/2}$ configuration.
Because of the high excitation energies, the gamma-ray spectrum is distributed in a wide range of energies, as shown in the top panel of Fig.~\ref{fig3}.
There are no significant peaks in the region not shown in the figure, with the integrated branching ratio being 2.71\% for $E_\gamma > $ 10 MeV.
The gamma-ray spectrum in the experimentally feasible range ($5~{\rm MeV} \leq E_\gamma \leq 9~{\rm MeV}$) is shown in the bottom panel of Fig.~\ref{fig3}.
Since there is no important discrete gamma-rays in this region (see Table \ref{tab2}), the branching ratio in this region is not large.
The integrated branching ratio between 5 and 9~MeV is 4.53\%.


\subsection{Two-proton decay in $^{16}$O}
\begin{figure}[t]
\begin{center}
\includegraphics[clip,width=7cm]{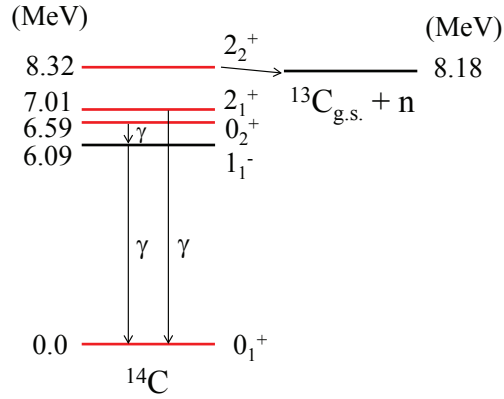}
\caption{A decay scheme for the excited states in $^{14}$C originated from the diproton decay from the 1$p_{1/2}$ and 1$p_{3/2}$ orbits in $^{16}$O.}
\label{fig4}
\end{center}
\end{figure}

Let us next discuss diproton decay from $^{16}$O, resulting in $^{14}$C.
The discussion is almost the same as that for the dineutron decay in the previous subsection.
Assuming the same configurations as in $^{14}$O, the populations of the ground state and the second 0$^+$ state at 6.59~MeV in $^{14}$C are 4.66\% and 10.9\%, respectively (see Table \ref{tab:Ppop14C}).
A big difference, however, is that the one neutron separation energy of $^{14}$C is 8.176~MeV, and the second 0$^+$ state decays to the first 1$^-$ state at 6.09~MeV by emitting a 0.50~MeV gamma-ray with a 100\% probability \cite{nndc}.
This state then decays to the ground state of $^{14}$C by emitting a 6.09~MeV gamma-ray (see Fig.~\ref{fig4}).
The first 2$^+$ state at 7.01 MeV, originating from an elimination of one proton from the 1$p_{1/2}$ level and one from the 1$p_{3/2}$ level, is also below the threshold for neutron emission, and this state decays to the ground state of $^{14}$C by emitting a 7.01 MeV gamma-ray.

\begin{table}[bht]
\caption{Same as Table \ref{tab:Ppop14O}, but for the daughter nucleus $^{14}$C for the $pp$ decay of $^{16}$O.
The numbers in the parentheses for the $\gamma$ decays are $\gamma$-ray energies, in units of MeV.}
\label{tab:Ppop14C}
\begin{center}
\begin{tabular}{cc|c|c|c|c}
\hline
\hline
$T$ & $I^\pi$ & configuration & $E^*$ (MeV) & $P_{\rm pop}$ & decay \\
\hline
1 & 0$^+$ & (1$p_{1/2}$)$^{-2}$ & 0 & 0.0466 & - \\
1 & 0$^+$ & (1$p_{3/2}$)$^{-2}$ & 6.59 & 0.109 & $\gamma$ (0.50+6.09) \\
1 & 2$^+$ & (1$p_{1/2}$)$^{-1}$(1$p_{3/2}$)$^{-1}$ & 7.01 & 0.201 & $\gamma$ (7.01)\\ 1 & 2$^+$ & (1$p_{3/2}$)$^{-2}$ & 8.32 & 0.109 & n emission\\
1 & 1$^-$ & (1$p_{1/2}$)$^{-1}$(1$s_{1/2}$)$^{-1}$ & B.W. & 0.119 & statistical\\
1 & 1$^-$ & (1$p_{3/2}$)$^{-1}$(1$s_{1/2}$)$^{-1}$ & B.W. & 0.270 & statistical \\
1 & 0$^+$ & (1$s_{1/2}$)$^{-2}$ & B.W. & 0.146 & statistical \\
\hline
\hline
\end{tabular}
\end{center}
\end{table}

\begin{figure}[t]
\begin{center}
\includegraphics[clip,width=8cm]{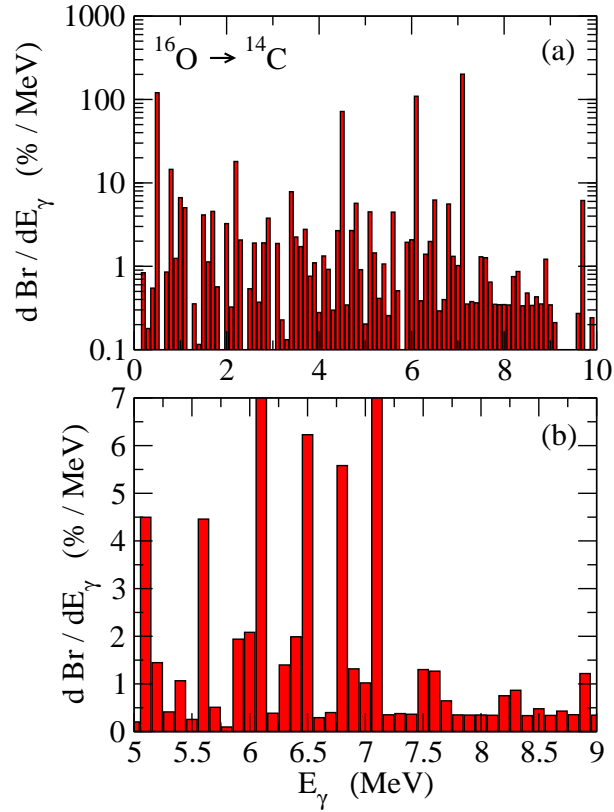}
\caption{
Same as Fig.~\ref{fig3}, but for the diproton decays of $^{16}$O.
The vertical axis of the lower panel is truncated at 7\% for presentation purposes, whereas the peaks at $E_\gamma=6.1$ and 7.1~MeV are as large as 109.3 and 201.6~\%/(0.1 MeV), respectively.
The accumulated branching ratio in the region not shown in the upper panel amounts to 2.36\%.}
\label{fig5}
\end{center}
\end{figure}

We estimate the branching ratios associated with the diproton decay from the $1s_{1/2}$ configuration using the {\tt TALYS} code as in the dineutron decay discussed in the previous subsection.
To this end, we use the mean excitation energy of $E^*_{0I}=-\epsilon_p(jl)-\epsilon_p(j'l')-S_{2p}(^{16}$O), where $S_{2p}(^{16}{\rm O})=22.33$~MeV is the two-proton separation energy of $^{16}$O.
We use the width of $\Gamma=7$~MeV as in the previous subsection.
The results are shown in Fig.~\ref{fig5}, Tables \ref{tab3} and \ref{tab4}, where we have also included the contribution of diproton decays from the $1p_{1/2}$ and the $1p_{3/2}$ configurations.
As one can see from the figures and tables, the gamma spectrum is dominated by the 7.01 MeV gamma-ray from the 2$_1^+$ state, as well as the 0.50~MeV and the 6.09~MeV gamma-rays originating from the population of the second 0$^+$ state.
Because of this, the integrated branching ratio between $E_\gamma=5$ and 9~MeV is now enhanced to 35.7\%.
Note that $E_\gamma=6.09$ and 7.01~MeV are in the favourable region for SNO+ experiment, which provides an ideal opportunity to search for invisible diproton decay of $^{16}$O.

\begin{table}[bht]
\caption{Same as Table \ref{tab1}, but for the diproton decays of $^{16}$O.}
\label{tab3}
\begin{center}
\begin{tabular}{cc|cc}
\hline
\hline
Nucleus & $\mathcal{B}$ (\%) & Nucleus & $\mathcal{B}$ (\%) \\
\hline
$^{14}$C (g.s.)  & 35.8 & $^{13}$C (g.s.) & 11.9 \\
$^{12}$C (g.s.)  & 13.8 & $^{12}$B (g.s.) & 1.13 \\
$^{11}$C (g.s.)  & 1.23 & $^{11}$B (g.s.) & 7.24 \\
$^{10}$B (g.s.)  & 2.04 & $^{10}$Be (g.s.) & 1.39 \\
$^{9}$Be (g.s.)  & 7.68 & $^{8}$Be (g.s.) & 16.07 \\
$^{7}$Li (g.s.)  & 2.65 & $^{6}$Li (g.s.) & 1.34 \\
\hline
\hline
\end{tabular}
\end{center}
\end{table}

\begin{table}[bht]
\caption{
Same as Table \ref{tab2}, but for the diproton decays of $^{16}$O.}
\label{tab4}
\begin{center}
\begin{tabular}{cccc}
\hline
\hline
Nucleus & Transition & $E_\gamma$ (MeV) & $\mathcal{B}$ (\%) \\
\hline
$^{14}$C & 0$_2^+$ (6.59) $\to$ 1$_1^-$ (6.09) & 0.50 & 10.9 \\
$^{14}$C & 1$_1^-$ (6.09) $\to$ 0$_1^+$ (0.0) & 6.09 & 10.9 \\
$^{14}$C & 2$_1^+$ (7.01) $\to$ 0$_1^+$ (0.0) & 7.01 & 20.1 \\
$^{12}$C & 2$_1^+$ (4.44) $\to$ 0$_1^+$ (0.0) & 4.44 & 6.56 \\
$^{11}$B & 1/2$_1^-$ (2.12) $\to$ 3/2$_1^-$ (0.0) & 2.12 & 1.50 \\
$^{10}$B & 1$_1^+$ (0.718) $\to$ 3$_1^+$ (0.0) & 0.718 & 1.33 \\
$^{7}$Li & 1/2$_1^-$ (0.478) $\to$ 3/2$_1^-$ (0.0) & 0.478 & 0.852 \\
\hline
\hline
\end{tabular}
\end{center}
\end{table}


\subsection{Proton-neutron decay in $^{16}$O}
\begin{figure}[t]
\begin{center}
\includegraphics[clip,width=7cm]{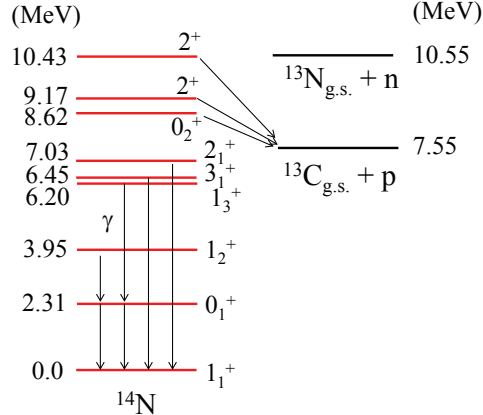}
\caption{A decay scheme for the excited states in $^{14}$N originated from the $pn$ decay from the 1$p_{1/2}$ and 1$p_{3/2}$ orbits in $^{16}$O.}
\label{fig6}
\end{center}
\end{figure}

We next consider $pn$ decay in $^{16}$O, resulting in $^{14}$N.
The population probability for each final state is summarized in Table \ref{tab:Ppop14N}.
In contrast to same-particle pairs (that is, $nn$ and $pp$), there are two possible proton-neutron combinations, that is, the isospin-singlet ($T=0$) and the isospin-triplet ($T$=1) configurations.
In the naive shell model, an elimination of the $pn$ pair from the 1$p_{1/2}$ state results in the ground state of $^{14}$N with $I^\pi=1^+$ and the first 0$^+$ state at 2.31 MeV, which are the isospin $T$=0 and 1 states, respectively.
The first 0$^+$ state decays to the ground state by emitting a 2.31~MeV gamma-ray \cite{nndc} as shown in Fig.~\ref{fig6}.
An elimination of a $pn$ pair from the 1$p_{3/2}$ state, on the other hand, results in a population of final states with $(T,I)=(1,0), (0,1), (1,2)$, and (0,3).
Among these, the $I^\pi=1^+$ state at 6.20 MeV and the 3$^+$ state at 6.45 MeV leads to the gamma-decays, while the 0$^+$ state at 8.62 MeV and the 2$^+$ state at 10.43 MeV decay by emitting a proton to the ground state of $^{13}$C \cite{nndc} (see Fig.~\ref{fig6} and Table \ref{tab:Ppop14N}).
In addition, the $I^\pi=1^+$ state at 3.95 MeV and the 2$^+$ state at 7.03 MeV originated from an elimination of two protons from the $1p_{1/2}$ and the $1p_{3/2}$ levels also lead to the gamma-ray emissions.

\begin{table}[bht]
\caption{Same as Tables \ref{tab:Ppop14O} and \ref{tab:Ppop14C}, but for  the daughter nucleus $^{14}$N for the $pn$ decay of $^{16}$O.}
\label{tab:Ppop14N}
\begin{center}
\begin{tabular}{cc|c|c|c|c}
\hline
\hline
$T$ & $I^\pi$ & configuration & $E^*$ (MeV) & $P_{\rm pop}$ & decay \\
\hline
0 & 1$^+$ & (1$p_{1/2}$)$^{-2}$ & 0 & 0.0413 & - \\
1 & 0$^+$ & (1$p_{1/2}$)$^{-2}$ & 2.31 & 0.0138 & $\gamma$ (2.31) \\
0 & 1$^+$ & (1$p_{1/2}$)$^{-1}$(1$p_{3/2}$)$^{-1}$ & 3.95 & 0.0890 & $\gamma$ (1.64+2.31)\\
0 & 1$^+$ & (1$p_{3/2}$)$^{-2}$ & 6.20 & 0.0580 & $\gamma$ (3.89+2.31) \\
0 & 3$^+$ & (1$p_{3/2}$)$^{-2}$ & 6.45 & 0.0773 & $\gamma$ (6.45) \\
0 & 2$^+$ & (1$p_{1/2}$)$^{-1}$(1$p_{3/2}$)$^{-1}$ & 7.03 & 0.0890 & $\gamma$ (7.03) \\
1 & 0$^+$ & (1$p_{3/2}$)$^{-2}$ & 8.62 & 0.0322 & p emission\\
1 & 2$^+$ & (1$p_{1/2}$)$^{-1}$(1$p_{3/2}$)$^{-1}$ & 9.17 & 0.0593 & p emission\\
1 & 2$^+$ & (1$p_{3/2}$)$^{-2}$ & 10.43 & 0.0322 & p emission\\
0 & 0$^-$ & (1$p_{1/2}$)$^{-1}$(1$s_{1/2}$)$^{-1}$ & B.W. & 0.0350 & statistical \\
0 & 1$^-$ & (1$p_{1/2}$)$^{-1}$(1$s_{1/2}$)$^{-1}$ & B.W. & 0.0700 & statistical \\
1 & 1$^-$ & (1$p_{1/2}$)$^{-1}$(1$s_{1/2}$)$^{-1}$ & B.W. & 0.0350 & statistical \\
0 & 1$^-$ & (1$p_{3/2}$)$^{-1}$(1$s_{1/2}$)$^{-1}$ & B.W. & 0.0398 & statistical \\
0 & 2$^-$ & (1$p_{3/2}$)$^{-1}$(1$s_{1/2}$)$^{-1}$ & B.W. & 0.119 & statistical \\
1 & 1$^-$ & (1$p_{3/2}$)$^{-1}$(1$s_{1/2}$)$^{-1}$ & B.W. & 0.0796 & statistical \\
0 & 1$^+$ & (1$s_{1/2}$)$^{-2}$ & B.W. & 0.0860 & statistical \\
1 & 0$^+$ & (1$s_{1/2}$)$^{-2}$ & B.W. & 0.0430 & statistical\\
\hline
\hline
\end{tabular}
\end{center}
\end{table}

\begin{figure}[t]
\begin{center}
\includegraphics[clip,width=8cm]{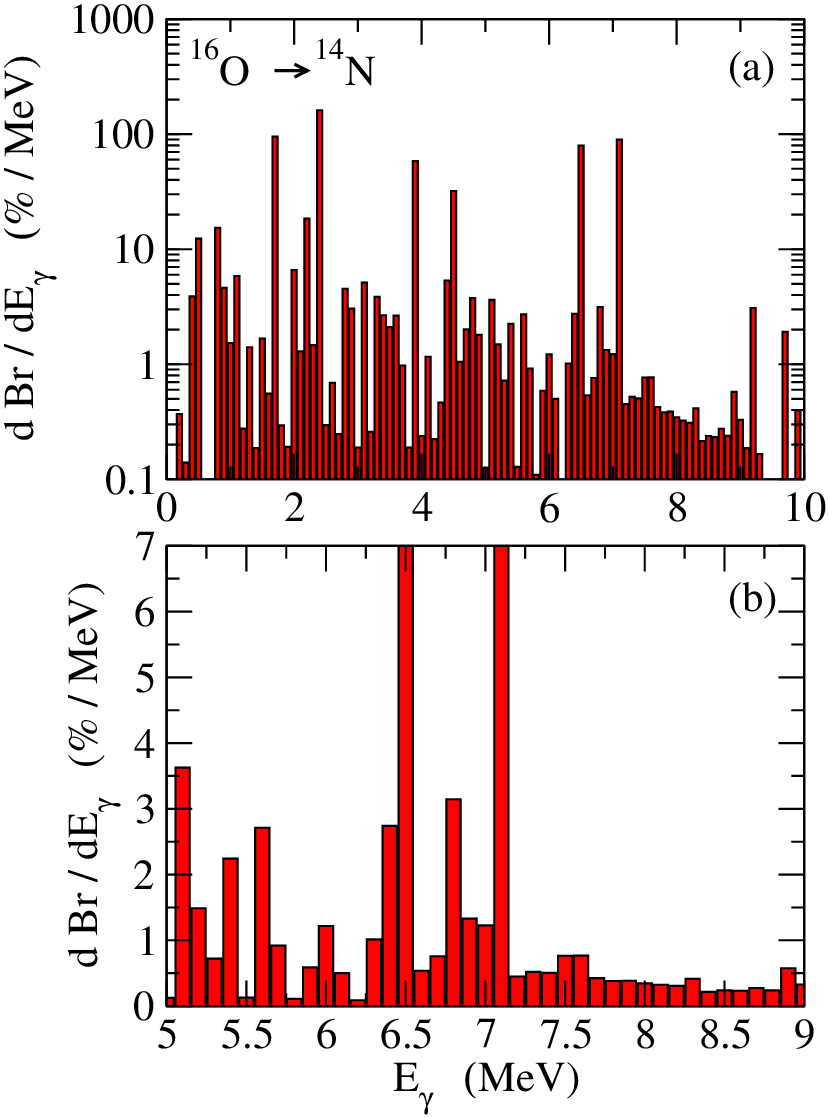}
\caption{
Same as Figs.~\ref{fig3} and \ref{fig5}, but for the $pn$ decays of $^{16}$O.
The vertical axis of the lower panel is truncated at 7\% for presentation purposes, whereas the peaks at $E_\gamma=6.5$ and 7.1~MeV are as large as 79.5 and 89.7~\%/(0.1 MeV), respectively.
The accumulated branching ratio in the region not shown in the upper panel amounts to 1.5\%.}
\label{fig7}
\end{center}
\end{figure}

The branching ratios associated with the $pn$ decay from the 1$s_{1/2}$ configuration are evaluated with the {\tt TALYS} code using the width of $\Gamma=7$~MeV.
The mean excitation energies are estimated by taking an average of the single particle energies for neutron and proton as $E^*_{0I}=-[\epsilon_p(jl)+\epsilon_n(j'l')+\epsilon_p(j'l')+\epsilon_n(jl)]/2 -S_{pn}(^{16}$O$)$, where $S_{pn}(^{16}$O) = 22.96 MeV is the energy required to remove one proton and one neutron from $^{16}$O.
The results are shown in Fig.~\ref{fig7}, Tables \ref{tab5} and \ref{tab6}, where we have added the contribution of $pn$ decays from the $1p_{1/2}$ and the $1p_{3/2}$ configurations.
One can see that the gamma-ray spectrum is dominated by the discrete gamma-rays, which originate from the populations of the discrete states in $^{14}$N below the threshold for proton emission, as well as the 2$^+$ state in $^8$Be.
In particular, the 6.45 MeV and 7.03 MeV gamma-rays have branching ratios of 7.73\% and 8.90\%, respectively (see Table \ref{tab6}).
The accumulated branching ratios in the region 5~MeV~$\leq E_\gamma \leq 9$~MeV 
amount to 20.2\%.
While this value is not as large as the corresponding branching ratios for the $pp$ decay, it is still large enough to be amenable to current experiments.

\begin{table}[bht]
\caption{Same as Tables \ref{tab1} and \ref{tab3}, but for the $pn$ decays of $^{16}$O.}
\label{tab5}
\begin{center}
\begin{tabular}{cc|cc}
\hline
\hline
Nucleus & $\mathcal{B}$ (\%) & Nucleus & $\mathcal{B}$ (\%) \\
\hline
$^{14}$N (g.s.) & 36.8 & $^{13}$C (g.s.) & 12.7 \\
$^{12}$C (g.s.) & 5.22 & $^{11}$C (g.s.) & 2.61 \\
$^{11}$B (g.s.) & 5.73 & $^{10}$B (g.s.) & 2.29 \\
$^{9}$B (g.s.) & 2.23 & $^{9}$Be (g.s.) & 4.50 \\
$^{8}$Be (g.s.) & 22.5 & $^{7}$Be (g.s.) & 0.97 \\
$^{7}$Li (g.s.) & 1.68 & $^{6}$Li (g.s.) & 2.41 \\
\hline
\hline
\end{tabular}
\end{center}
\end{table}

\begin{table}[bht]
\caption{
Same as Tables \ref{tab2} and \ref{tab4}, but for the $pn$ decays of $^{16}$O.}
\label{tab6}
\begin{center}
\begin{tabular}{cccc}
\hline
\hline
Nucleus & Transition & $E_\gamma$ (MeV) & $\mathcal{B}$ (\%) \\
\hline
$^{14}$N & 1$_2^+$ (3.95) $\to$ 0$_1^+$ (2.31) & 1.64 & 8.9 \\
$^{14}$N & 0$_1^+$ (2.31) $\to$ 1$_1^+$ (0.0) & 2.31 & 16.1 \\
$^{14}$N & 1$_3^+$ (6.20) $\to$ 0$_1^+$ (2.31) & 3.89 & 5.80 \\
$^{14}$N & 3$_1^+$ (6.45) $\to$ 1$_1^+$ (0.0) & 6.45 & 7.73 \\
$^{14}$N & 2$_1^+$ (7.03) $\to$ 1$_1^+$ (0.0) & 7.03 & 8.90 \\
$^{12}$C & 2$_1^+$ (4.44) $\to$ 1$_1^+$ (0.0) & 4.44 & 2.71 \\
$^{11}$B & 1/2$_1^-$ (2.12) $\to$ 3/2$_1^-$ (0.0) & 2.12 & 1.27 \\
$^{10}$B & 1$_1^+$ (0.718) $\to$ 3$_1^+$ (0.0) & 0.718 & 1.48 \\
\hline
\hline
\end{tabular}
\end{center}
\end{table}


\section{Summary}
We have evaluated the branching ratios associated with baryon number non-conserving dinucleon decays in $^{16}$O.
In particular, we have investigated the gamma-spectra in the experimentally relevant energy region between 5 and 9~MeV.
For the decays from the 1$p_{1/2}$ and the 1$p_{3/2}$ configurations in $^{16}$O, we took advantage of the known decay properties of the daughter nuclei, while for the decay from the 1$s_{1/2}$ configuration we used the statistical model provided by the {\tt TALYS} software.

For $nn$ decay in $^{16}$O, we did not find appreciable branching ratios for gamma-rays in the region 5~MeV~$\leq E_\gamma \leq 9$~MeV.
In contrast, for $pp$ decay, we found that the discrete gamma-rays with energies 6.09~MeV and 7.01 MeV have significant branching ratios of 10.9\% and 20.1\%, respectively.
For $pn$ decay, gamma-rays with energies 6.45~MeV and 7.03 MeV have branching ratios as high as 7.73\% and 8.9\%, respectively.
These gamma-rays are within the favourable energy region for the initial water phase of the SNO+ experiment, and provide a promising way to search for 
dinucleon decay in $^{16}$O.

The branching ratios evaluated in this paper, together with the branching ratios for the single-nucleon decays shown in Refs.~\cite{Ejiri93,Kamyshkov03}, will be necessary ingredients in evaluating the lower limit of the invisible nucleon decays.
We expect that our results will be useful in both current and future experiments, such as SNO+.


\ack
We thank H.~Ejiri and Y.~Tanimura for useful discussions.
This work was supported by the Science and Technology Facilities 
Council [STFC grant ST/N000307/1].


\appendix
\section{Population probabilities for the final state of daughter nuclei} 
In this Appendix, we give a detailed derivation of Eqs.~(\ref{popP0}) and (\ref{popP1}) describing the population probabilities for the final states of the daughter nuclei produced via dinucleon decay in $^{16}$O.


\subsection{$nn$ and $pp$ decays} 
We first discuss the $nn$ and $pp$ decays.
Both of these decays can be treated in the same way, and we consider only the $nn$ decay here.
In this case, the isospin quantum numbers in Eq.~(\ref{pop_prob}) are restricted to $t_z=t_z'=n$ and the isospin in the final state, Eq.~(\ref{finalwf}), is $T=1$ and $T_z=-1$.
The population probability thus reads, %
\begin{equation}
P_{\rm pop}(f)\propto
\sum_{s_z,s_z'}\int d\vec{r}
\left|\left\langle jlj'l';IM\left|a_{\vec{r}s_zn}a_{\vec{r}s'_zn}\right
|^{16}{\rm O}\right\rangle\right|^2, \label{Ppop0-appen}
\end{equation}
with %
\begin{equation}
|jlj'l'; IM\rangle
=
{\cal N}\sum_{m,m'}
\langle jmj'm'|IM\rangle
a_{jlmn}a_{j'l'm'n}|^{16}{\rm O}\rangle.
\end{equation}
Notice that the nucleon creation operator $a^\dagger_{\vec{r}s_zn}$ can be expanded as \cite{Dobaczewski84}, %
\begin{equation}
a^\dagger_{\vec{r}s_zn}=\sum_{j,l,m}
\langle \chi_{s_z}|\psi^{(n)}_{jlm}(\vec{r})\rangle a_{jlmn}^\dagger, \end{equation}
where $\chi_{s_z}$ is the spin wave function and $\psi^{(n)}_{jlm}(\vec{r})$ is the neutron single-particle wave function given by %
\begin{equation}
\psi^{(n)}_{jlm}(\vec{r})=\phi^{(n)}_{jl}(r){\cal Y}_{jlm}(\hat{\vec{r}}).
\end{equation}
Here, $\phi^{(n)}_{jl}(r)$ is the radial wave function, and ${\cal Y}_{jlm}(\hat{\vec{r}})$ is the spin-angular wave function given by %
\begin{equation}
{\cal Y}_{jlm}(\hat{\vec{r}})
=\sum_{m_l,m_s}\left.\left\langle l m_l \frac{1}{2} m_s\right|j m\right\rangle\,Y_{lm_l}(\hat{\vec{r}})\chi_{m_s}, \end{equation}
where $Y_{lm_l}(\hat{\vec{r}})$ is the spherical harmonic function.
Therefore, the matrix element in Eq.~(\ref{Ppop0-appen}) reads, 
\begin{eqnarray}
&&\left\langle ^{16}{\rm O}\left|a^\dagger_{\vec{r}s'_zn}
a^\dagger_{\vec{r}s_zn}\right|
jlj'l';IM\right\rangle \nonumber \\
&=&
{\cal N}\sum_{m,m'}
\langle jmj'm'|IM\rangle 
\sum_{\tilde{j},\tilde{l},\tilde{m}}\sum_{\tilde{j}',\tilde{l}',\tilde{m}'}
\langle \chi_{s'_z}|\psi^{(n)}_{\tilde{j}'\tilde{l}'\tilde{m}'}(\vec{r})
\rangle \langle \chi_{s_z}|\psi^{(n)}_{\tilde{j}\tilde{l}\tilde{m}}(\vec{r})
\rangle \nonumber \\
&&\times
\langle ^{16}{\rm O}|a_{\tilde{j}'\tilde{l}'\tilde{m}'n}^\dagger
a_{\tilde{j}\tilde{l}\tilde{m}n}^\dagger
a_{jlmn}a_{j'l'm'n}| ^{16}{\rm O}\rangle.
\end{eqnarray}
Notice the relation 
\begin{equation}
a_{jlmn}^\dagger|^{16}{\rm O}\rangle=0, \end{equation}
for 1$s_{1/2}$, 1$p_{3/2}$, and 1$p_{1/2}$.
Using Wick's theorem, one thus obtains the following relation: 
\begin{eqnarray}
&&\langle ^{16}{\rm O}|a_{\tilde{j}'\tilde{l}'\tilde{m}'n}^\dagger
a_{\tilde{j}\tilde{l}\tilde{m}n}^\dagger
a_{jlmn}a_{j'l'm'n}| ^{16}{\rm O}\rangle \nonumber \\
&&=
\delta_{jlm,\tilde{j}\tilde{l}\tilde{m}}
\delta_{j'l'm',\tilde{j}'\tilde{l}'\tilde{m}'}
-\delta_{jlm,\tilde{j}'\tilde{l}'\tilde{m}'}
\delta_{j'l'm',\tilde{j}\tilde{l}\tilde{m}}.
\end{eqnarray}
With this relation, one obtains, %
\begin{eqnarray}
&&\left\langle ^{16}{\rm O}\left|a^\dagger_{\vec{r}s'_zn}
a^\dagger_{\vec{r}s_zn}\right|
jlj'l';IM\right\rangle  \nonumber \\
&=&{\cal N}\sum_{m,m'} 
\left\{
(-)^{j+j'-I}\langle j'm'jm|IM\rangle \right. 
\langle \chi_{s'_z}|\psi^{(n)}_{j'l'm'}(\vec{r})\rangle 
\langle \chi_{s_z}|\psi^{(n)}_{jlm}(\vec{r})\rangle \nonumber \\
&&\hspace*{-0.5cm}\left.-
\langle jmj'm'|IM\rangle \langle \chi_{s'_z}|\psi^{(n)}_{jlm}(\vec{r})\rangle \langle \chi_{s_z}|\psi^{(n)}_{j'l'm'}(\vec{r})\rangle \right\}.
\end{eqnarray}

Furthermore, we use the helicity representation for the spin-angular wave function \cite{BE91}, %
\begin{equation}
{\cal Y}_{jlm}(\hat{\vec{r}})
=\sum_h \frac{\hat{j}}{\sqrt{8\pi}}\,D^j_{mh}(\hat{\vec{r}})\, \alpha_{jlh}\chi_h, \end{equation}
where $D^j_{mh}(\hat{\vec{r}})$ is the Wigner's $D$-function, $\hat{j}$ is defined as $\hat{j}\equiv\sqrt{2j+1}$, and $\alpha_{jlh}$ is given as $\alpha_{jlh}=(-)^{(h+1/2)(j-l-1/2)}$.
Using the relation for the $D$-functions, 
\begin{eqnarray}
&&\sum_{M_1,M_2}\langle I_1M_1I_2M_2|IM\rangle D^{I_1}_{M_1M_1'}(\hat{\vec{r}})D^{I_2}_{M_2M_2'}(\hat{\vec{r}}) 
=\langle I_1M_1'I_2M_2'|IM'\rangle D^{I}_{MM'}(\hat{\vec{r}}), \nonumber \\
\end{eqnarray}
one obtains 
\begin{eqnarray}
&&\left\langle ^{16}{\rm O}\left|a^\dagger_{\vec{r}s'_zn}
a^\dagger_{\vec{r}s_zn}\right|
jlj'l';IM\right\rangle 
=
{\cal N}\phi_{jl}^{(n)}(r)\phi_{j'l'}^{(n)}(r)\, \frac{\hat{j}\hat{j}'}{\sqrt{8\pi}}\,D^I_{M s_z+s_z'}(\hat{\vec{r}}) \nonumber \\
&&\times
\left\{
(-)^{j+j'-I}\langle j's_z'js_z|Is_z+s_z'\rangle \alpha_{jls_z}\alpha_{j'l's'_z} 
-\langle js_z'j's_z|Is_z+s_z'\rangle \alpha_{jls'_z}\alpha_{j'l's_z}
\right\}. \nonumber \\
\end{eqnarray}
This quantity vanishes when $s_z=s_z'$.
One thus obtains %
\begin{eqnarray}
&&\left\langle ^{16}{\rm O}\left|a^\dagger_{\vec{r}s'_zn}
a^\dagger_{\vec{r}s_zn}\right|
jlj'l';IM\right\rangle 
=
{\cal N}\phi_{jl}^{(n)}(r)\phi_{j'l'}^{(n)}(r)\, \frac{\hat{j}\hat{j}'}{\sqrt{8\pi}}\,D^I_{M 0}(\hat{\vec{r}})\, \langle js_zj's'_z|I0\rangle \nonumber \\
&&\times
\left\{
\alpha_{jls_z}\alpha_{j'l's'_z} -(-)^{j+j'-I}
\alpha_{jls'_z}\alpha_{j'l's_z}
\right\}.
\end{eqnarray}
Notice that this is the same as Eq.~(A.11) in Ref.~\cite{BE91} except for an overall factor and a phase.
Using the orthogonality relation of the $D$ function, %
\begin{equation}
\int d\hat{\vec{r}}\,D^I_{MK}(\hat{\vec{r}})^*
D^{I'}_{M'K'}(\hat{\vec{r}})=\frac{4\pi}{2I+1}\,\delta_{I,I'}\delta_{M,M'}\,\delta_{K,K'}, \end{equation}
one finally obtains Eq.~(\ref{popP1}) with %
\begin{equation}
I_r=\int r^2dr\,\left[\phi^{(n)}_{jl}(r)\right]^2\left[\phi^{(n)}_{j'l'}(r)\right]^2, \label{I_r}
\end{equation}
where we have used the relation %
\begin{eqnarray}
&& \sum_{s_z,s_z'}\langle js_zj's'_z|I0\rangle^2 
\left\{
\alpha_{jls_z}\alpha_{j'l's'_z} -(-)^{j+j'-I}
\alpha_{jls'_z}\alpha_{j'l's_z}
\right\}^2 \nonumber \\
&&=2 \left.\left\langle j\frac{1}{2}j'-\frac{1}{2}\right|I0\right\rangle^2\, (1+(-1)^{l+l'-I})^2.
\end{eqnarray}


\subsection{$pn$ decay}
Let us now discuss the $pn$ decay.
In this case, the isospin quantum numbers in Eq.~(\ref{pop_prob}) are restricted to $t_z=-t_z'$ and the $z$-component of the isospin in the final state is $T_z=0$.
The isospin in the final state can be either $T=1$ or $T=0$.

Following the same procedure as in the $nn$ decay, one obtains, %
\begin{eqnarray}
&&
\left\langle ^{16}{\rm O}\left|a^\dagger_{\vec{r}s'_zn}
a^\dagger_{\vec{r}s_zp}\right|
jlj'l';IM\right\rangle \nonumber \\
&=&
{\cal N}\phi_{jl}(r)\phi_{j'l'}(r)\, \frac{\hat{j}\hat{j}'}{\sqrt{8\pi}}\,D^I_{M s_z+s_z'}(\hat{\vec{r}}) \left.\left\langle \frac{1}{2}-\frac{1}{2}\frac{1}{2}\frac{1}{2}\right|T0\right\rangle
\nonumber \\
&&\times
\left\{
(-)^{j+j'-I+1-T}\langle j's_z'js_z|Is_z+s_z'\rangle \alpha_{jls_z}\alpha_{j'l's'_z} \right. \nonumber \\
&&\left. -
\langle js_z'j's_z|Is_z+s_z'\rangle \alpha_{jls'_z}\alpha_{j'l's_z}
\right\}, \end{eqnarray}
and a similar expression for $\left\langle ^{16}{\rm O}\left|a^\dagger_{\vec{r}s'_zp}
a^\dagger_{\vec{r}s_zn}\right|
jlj'l';IM\right\rangle$.
Here, we have assumed that the single-particle wave function is the same between proton and neutron for a given $j$ and $l$, that is, $\phi_{jl}^{(n)}(r)=\phi_{jl}^{(p)}(r)\equiv \phi_{jl}(r)$.

From this expression, one obtains the same population probability for $T=1$ as in the $nn$ and $pp$ decays, Eq.~(\ref{popP1}).
For $T=0$, by using the following relations, %
\begin{eqnarray}
&&\left\{
(-)^{j+j'-I+1}\left.\left\langle j'\pm\frac{1}{2}j\pm\frac{1}{2}
\right|I\pm1\right\rangle \alpha_{jl\pm1/2}\alpha_{j'l'\pm1/2} \right. \nonumber \\
&&
\left. -
\left.\left\langle j\pm\frac{1}{2}j'\pm\frac{1}{2}\right|I\pm1\right\rangle \alpha_{jl\pm1/2}\alpha_{j'l'\pm1/2} \right\}^2 
=4
\left.\left\langle j\frac{1}{2}j'\frac{1}{2}\right|I1\right\rangle ^2, \\
&&\left\{
(-)^{j+j'-I+1}\left.\left\langle j'\mp\frac{1}{2}j\pm\frac{1}{2}\right|I0\right\rangle \alpha_{jl\pm1/2}\alpha_{j'l'\mp1/2} \right. \nonumber \\
&&
\left. -
\left.\left\langle j\mp\frac{1}{2}j'\pm\frac{1}{2}\right|I0\right\rangle \alpha_{jl\mp1/2}\alpha_{j'l'\pm1/2} \right\}^2 
=
\left.\left\langle j\frac{1}{2}j'-\frac{1}{2}\right|I0\right\rangle ^2\, (1-(-1)^{l+l'-I})^2. \nonumber \\
\end{eqnarray}
one finally obtains Eq.~(\ref{popP0}).



\end{document}